# Mathematical Models in Danube Water Quality

**Valerian Antohe, Constantin Stanciu**
**Faculty of Engineering, Brăila, România**

**ABSTRACT**: The mathematical shaping in the study of water quality has become a branch of environmental engineering. The comprehension and effective application of mathematical models in studying environmental phenomena keep up with the results in the domain of mathematics and the development of specialized software as well. Integrated software programs simulate and predict extreme events, propose solutions, analyzing and processing data in due time. This paper presents a browsing through some mathematical categories of processing the statistical data, examples and their analysis concerning the degree of water pollution downstream the river Danube.

**Introduction**

Principles of mathematical shaping we live in a world of models. The consciousnesses of the ego are simplification of our human being meaning models. Each one's life is a continuous confrontation between the shape of the ego and the model of the surrounding world. All the science were born and significantly evolved after models had been elaborated [Neu03].

The paradigms of shaping a certain process acknowledge that an overlook upon similar models created for the issue under discussion is needed. Sometimes an adjustment or an optimization of already existing models can be reached; gathering the useful information is necessary in order to understand all the parameters of the issue and initial designing of a simplified model which can be further improved leads to achieving a flexible model. It is also necessary that all the extreme cases should be analyzed and a ranking of results should be created as well as sorting the information according to the degree of the impact with the expectancy and the results estimation. There are rarely perfect solutions. Shaping is an art of discovery as well as of satisfying the compromise. A certain process is normally analyzed according to the statistic data. Finding the function which achieves the connection which achieves the evolution of this data is of a big importance.

## 1 Motivation

The general proprieties of natural waters are mainly determined in the solid, liquid and gas substances existing as suspension or diluted materials. The very numerous substances come from the interactions hydrosphere-atmosphere-lithosphere-live organism. thus, in a study conducted by Cousteau team (1991-1992) on the quality of the Danube's waters, it has been highlighted that are 800 organic and inorganic compounds, among which 50% are to be found in vegetal and animal tissues for underwater medium. Each type of source presents its own physical, chemical and biological characteristics, varying from a region to another according to the mineralogical composition of the areas it crosses, to the contact duration, to temperature and climatic conditions.

We assume that Danube water breathe like a living being. How can we control this process using statistical data about, predict, analyze, prevent the critical period, those are some goals for this review.

## 2 Mathematical model

We consider the prediction problem of continuous - time stochastic process on an entire time-interval in terms of its recent past. In the deterministic literature, such problems are usually solved by suitable regularization techniques [AS03]. We try to obtain a fast and feasible prediction using some statistic data for dissolved oxygen (fig.3.2.). Statistical data correspond to table of determination on three places nearly Braila, i.e. Gropeni Station, (fig.3.1), Braila 1 and Braila 2. Smoothing spline interpolation will be used, compared with Lagrange interpolation. We also compare the resulting predictions with those obtained by other methods available and establish the same methods of analyzing for different annual time or different places.

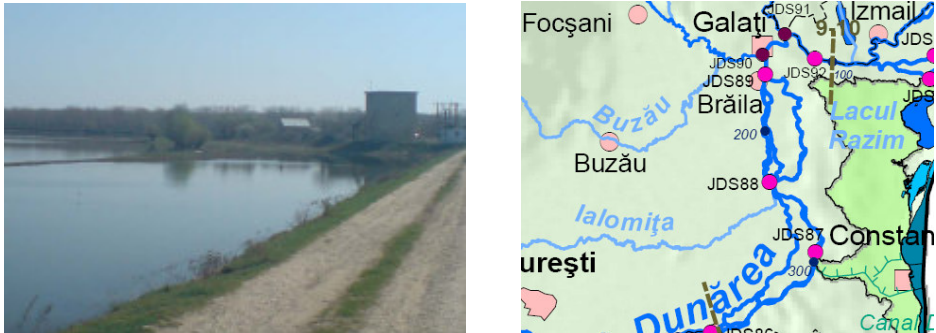

*Figure 3.1* The APM Station in Gropeni Braila and the map with three location for water quality data: Dunare-Gropeni, Dunare Braila-1, Dunare Braila-2

In the analyzed examples the possibility of numerical analysis of same data by means of cubic spline function is used.

| Data | temp. $^0$C | pH | OD | CBO5 | CCO-Mn | CCO-Cr |
|---|---|---|---|---|---|---|
| 9/11/2003 | 21 | 7.5 | 8.1 | 4.8 | 6.4 | 20 |
| 10/14/2003 | 14 | 7.5 | 7.5 | 6.8 | 16.8 | - |
| 11/11/2003 | 11 | 7.2 | 7.9 | 4.7 | 8.8 | 20 |
| 12/5/2003 | 7 | 7.5 | 7.3 | 5.9 | 9.6 | 20 |
| 1/30/2004 | 3 | 7.2 | 8.3 | 14 | 24 | 45 |
| 2/5/2004 | 3 | 6.8 | 8.5 | 7.7 | 13.6 | 30 |
| 3/25/2004 | 9 | 7.6 | 9.2 | 7 | 12.8 | 20 |
| 4/30/2004 | * | 7.1 | 9.8 | 5.8 | 8 | 30 |
| 5/31/2004 | 20 | 8 | 8 | 4.9 | 6.4 | 14.4 |
| 6/28/2004 | 26 | 7.9 | 9 | 5.5 | 8 | 19.2 |
| 7/15/2004 | * | 7.9 | 7.6 | 7.9 | 15.2 | 33.6 |

*Figure 3.2* Statistical Data Dunare-Gropeni

The term "interpolation" is understood as the determination of the value of an unknown function within the given interval. In order to make an interpolation the unknown function is approximated with an appropriate analytical function, the value needed being determined afterwards. The polynomial approach is most frequently used. The theoretical basis of polynomial approach is the Weierstrass theorem, in which it is shown that any continuous function can be approximated with an appropriate precision on a given close interval by a polynomial form. Unfortunately, Weierstrass theorem does not provide any practical criterion of finding the right polynomial form. In many cases the form of the function f can be "guessed".

Interpolation can be made through approaching with different polynomial form: with divided differences, with finite differences, or Lagrange polynomial form [Cli07]. The MATLAB soft provides a complete and unitary solution for approximating problems. This medium is used both by mathematicians and engineers, [Mur07].

Observing data at point $x_i$, i=1…12 (annual) or multiply with number of annual decades, $y_i$ will be the registered data for OD (for example). Then a cubic spline is a curve y=f(x) interpolating all points and consisting of a cubic polynomials between each consecutive pair of knots $x_i$ and $x_{i+1}$. The parameters of a cubic polynomials are constrained so that f(x) is continuous and it has continuous derivates f'(x) and f''(x), thus causing it to be smooth.

Usually, this is done by specifying the form of the function at two extremes, $x_1$ and $x_n$. A natural spline is obtained by requiring f''($x_1$)=f''($x_n$)=0, thus making the curve linear at the extremes (first data involved and last data). Those hypotheses are well done for our model, the concentration of any element observed being accepted without a sudden variation of monotony.

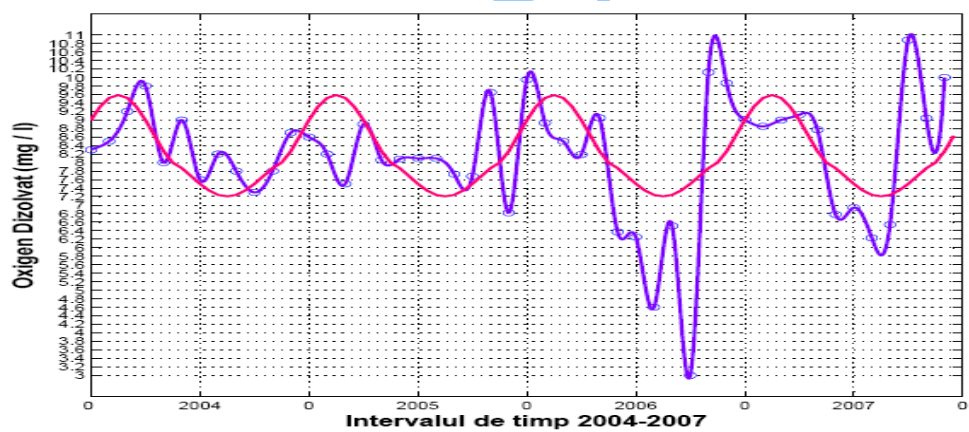

*Figure 3.3 Spline model for dissolved oxygen (blue)
and the harmonic function (red) 2004-2007
hr=(sin(8*pi*k1/192)+cos(8*pi*k1/192)).^(4/3)*

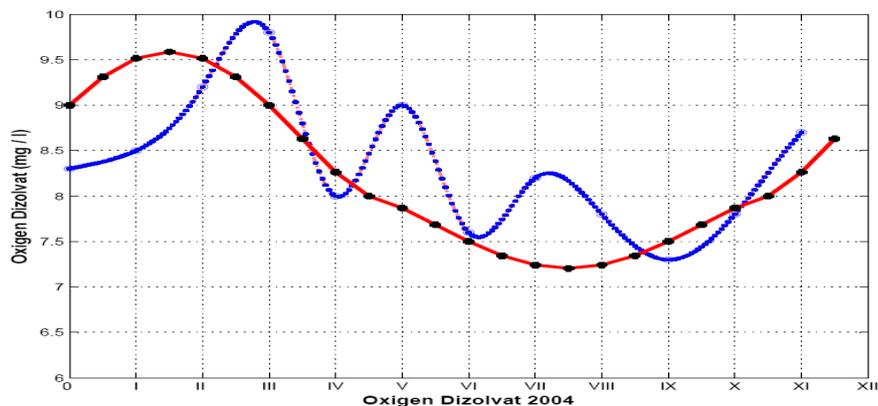

*Figure 3.4 Spline model for dissolved oxygen (blue) and the harmonic function (red), only for 2004.*

## Conclusions

Data analysis using polynomial regression show that there is a monotone decrease of DO during an year and most during four year, blue line, (fig.4.1, fig.4.2.).

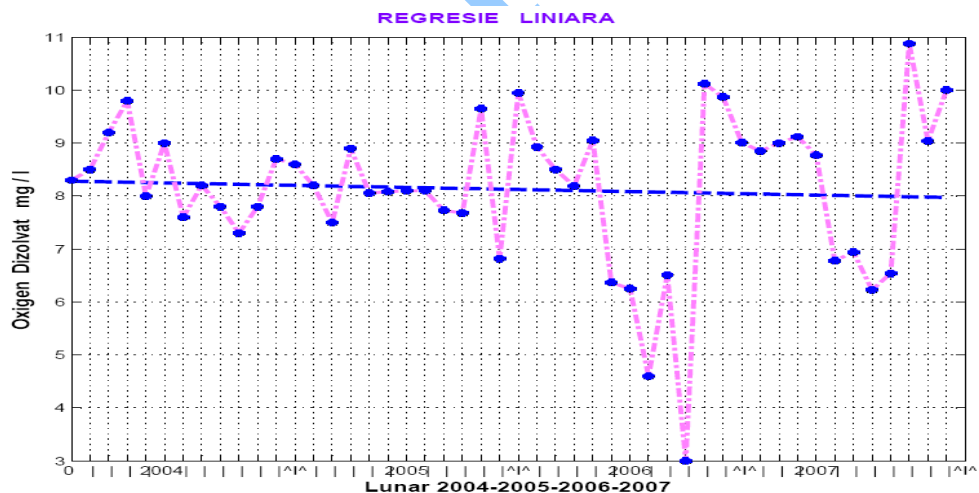

*Figure 4.1 linear regressions for 2004-2007 time intervals; show a down evolution of about 0.2 mg/l*

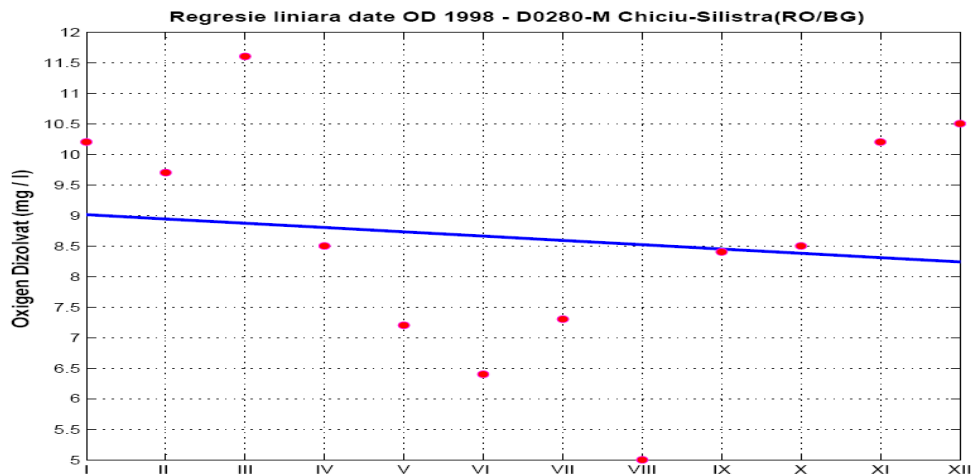

*Figure 4.2* Linear regression for only 1998 time interval, show a down evolution of about 0.3 mg/l OD (data registered on DO280-RO-BG-Station).

We consider that our results show a strong spatial structure for water characteristic on a given day and temporal structure from day to day, using a high degree of division, and some mathematical software, especially not in a predicted time interval (fig.4.3.), the red line representing the graphic of the function f(x) = (sinx+cosx)^(4/3), an inspired function in this case.

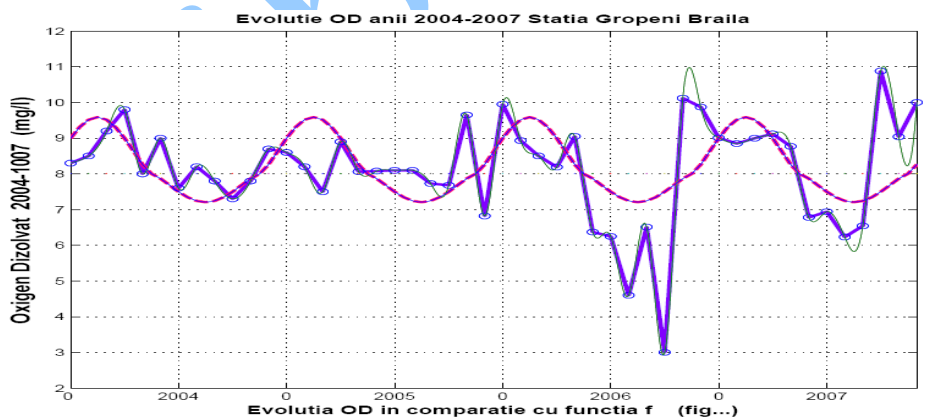

*Figure 4.3* Spline interpolation with a more fine norm show in the last decade of 2006 (for example) that dissolved oxygen has a grow up of about 1 mg/l (the red curve)

If a spline function will interpolate the data using a fine norm, as we see in fig. 4.4. and fig. 4.7., the graph indicate that between the second and the third month there is a maximum of 9.9 (mg/l), not visualized in data table. Many cases like this can be present during an annual decade so, we could accept the real evolution the red representing in graph.

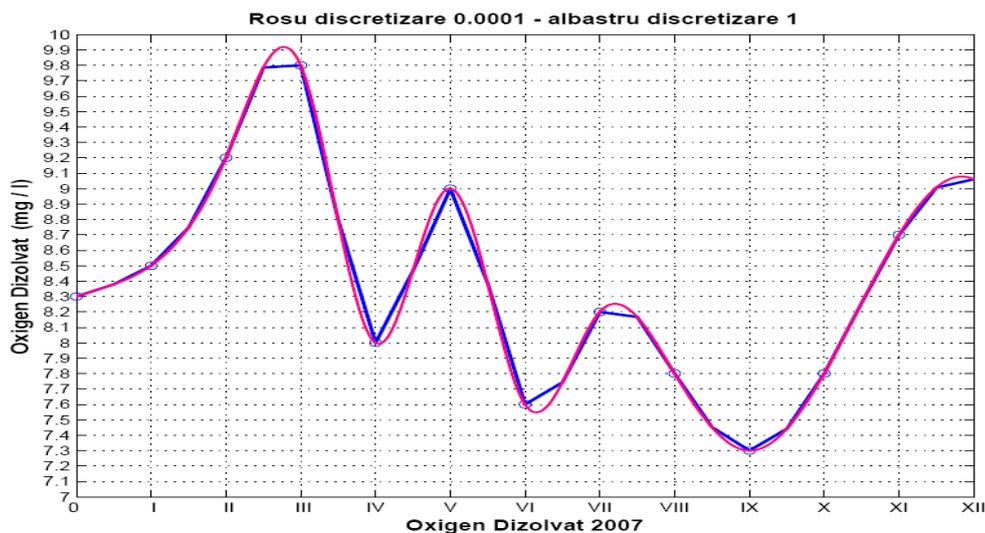

*Figure 4.4* *Spline interpolation with a more fine norm show some anomaly in 2007 (for example the 3-rd and the 6-th month)*

Understanding temporal variability of an element such is DO or the concordance with temperature evolution or other meteorological fields is an important way to scientific inquiry ecology, hydrology, environmental epidemiology, [M+08]. In some settings water pollution model may depend directly on prediction or interpolation of field such as rain, drought, floods. In others, model might built on and/or interact with others, rather than depend on prediction from that model, [IRF08].

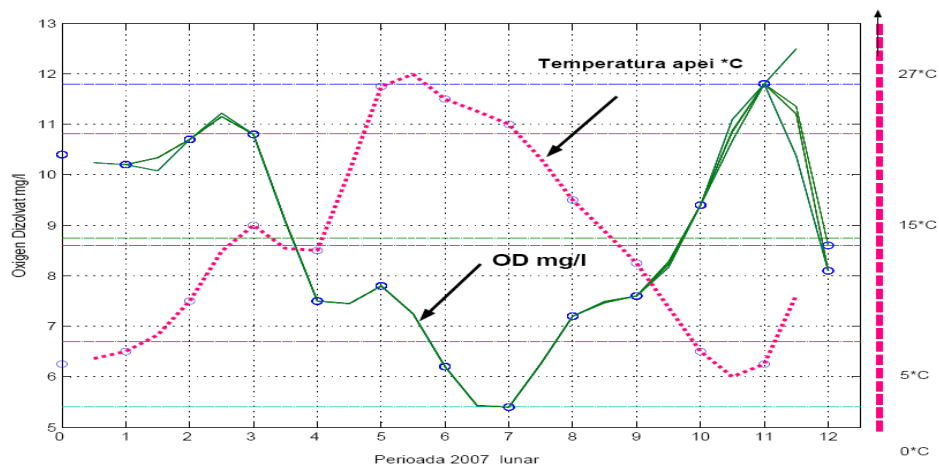

*Figure 4.5* Connection between water temperature
and the dissolved oxygen in 2007.

A water quality model usually consist of a set of mathematical expressions relating one or more water quality parameters, e.g., temperature, pH, conductivity, biochemical and sediment oxygen demand, heavy metals, organic and inorganic matter, etc., to one or natural processes. As we can see in fig. 4.5., there is a connection between water temperature and the dissolved oxygen and other parameters connection like in fig. 4.6.

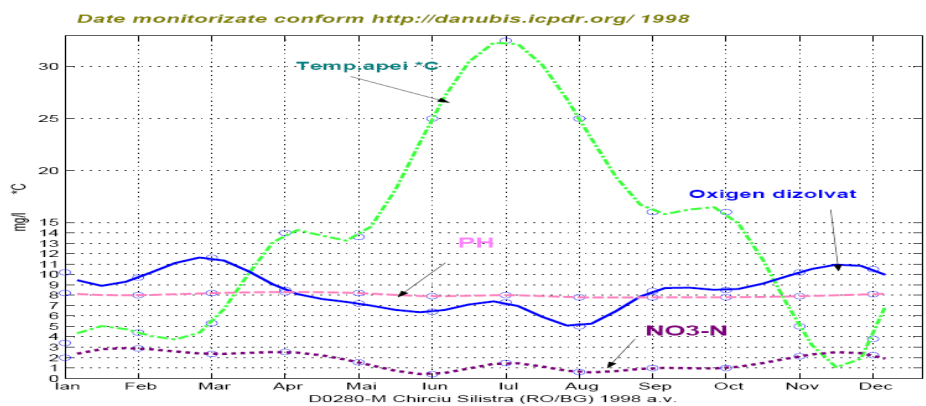

*Figure.4.6.* Monitorizated data in DO280-M Chiciu Silistra (RO/BG) conform icpdr.org expedition (water temperature, DO, Ph, NO3-N), during 1998.

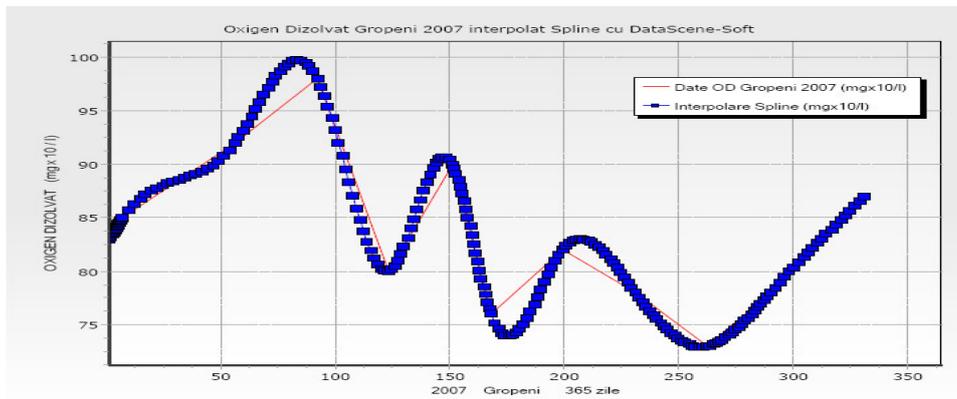

*Figure.4.7.* Data for Dissolved Oxygen interpolated with B-Spline, using DataScene-Soft.

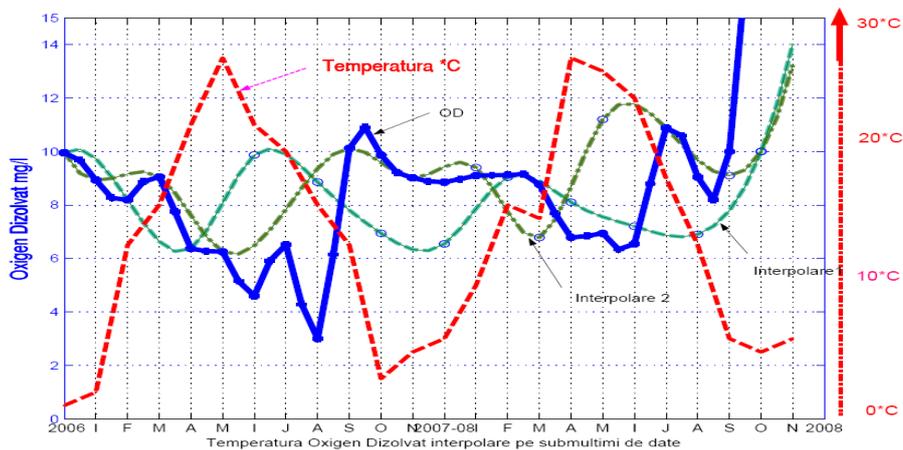

*Figure.4.8.* The green curve obtained for a subset of data, show only sometimes the real behavior of Dissolved Oxygen (the blue curve in the graph)

Future investigation will be focused to many other connections between water quality parameters, and finding some numerical function which can describe a good prediction.

45